\documentclass{PoS}

\usepackage{amsmath}
\usepackage{amsfonts}
\usepackage{amssymb}
\usepackage{bm}
\usepackage{array}
\usepackage{booktabs}
\usepackage{ulem}

\newcommand{\rep}[1]{\boldsymbol{#1}}
\newcommand{\brep}[1]{\overline{\boldsymbol {#1}}}
\newcommand{\eV}{\,\mathrm{eV}}
\newcommand{\MU}{M_{\mathrm U}}

\def\vev#1{\left\langle #1\right\rangle}
\def\abs#1{\left| #1\right|}

\def\Tr{\mbox{Tr}\,}

\def\eq#1{{Eq.~(\ref{#1})}}

\def\fig#1{{Fig.~\ref{#1}}}

\newcommand{\beq}{\begin{equation}}
\newcommand{\eeq}{\end{equation}}

\title{Axion poperties in GUTs}

\ShortTitle{Axion properties in GUTs}

\author{Anne Ernst,$^a$ Luca Di Luzio,$^{b}$
\speaker{Andreas Ringwald}$^{a}$
     and Carlos Tamarit$^{c}$\\
     \llap{$^a$}Deutsches Elektronen-Synchrotron DESY,
     Notkestr. 85, D-22607 Hamburg, University\\
     \llap{$^b$}Dipartimento di Fisica dell'Universit\`a di Pisa and INFN, Italy\\
     \llap{$^c$}Physik Department T70, Technische Universit\"{a}t M\"{u}nchen, 
     James Franck Stra\ss e 1, D-85748 Garching, Germany \\
     E-mail:  \email{anne.ernst@desy.de}, \email{luca.diluzio@pi.infn.it},
     \email{andreas.ringwald@desy.de}, \email{carlos.tamarit@tum.de}}

\abstract{We summarize recent studies of realistic nonsupersymmetric Grand Unified Theories (GUTs) extended with a global $U(1)_{\rm PQ}$ symmetry, so as to accommodate the axion solution to the strong CP problem. Aside from solving the CP problem and unifying the gauge structure of the SM, these models can also yield realistic spectra and mixings, including neutrino masses, and allowing for a consistent cosmological history that accounts for inflation, dark matter and baryogenesis. In our studies of  $SO(10)$ and $SU(5)$ theories, we determined the mass and couplings of the axion in terms of the relevant threshold scales, and assessed how the former are constrained from the requirements of gauge coupling unification, proton decay searches and collider bounds. The axion mass ends up being rather constrained for GUT scale axions, particularly in the case of $SU(5)$, and could be probed by upcoming dark matter experiments, such as  ABRACADABRA and CASPEr-Electric.
}

\FullConference{Corfu Summer Institute 2018 "School and Workshops on Elementary Particle Physics and Gravity"\\
		(CORFU2018)\\
		31 August - 28 September, 2018\\
		Corfu, Greece}

\begin{document}

\section{Introduction}
The Peccei-Quinn (PQ) solution to the strong CP problem \cite{Peccei:1977hh,Peccei:1977ur} has attracted much attention in recent times. It involves the postulation of additional scalar or fermionic degrees of freedom, constrained by a global anomalous $U(1)_{\rm PQ}$ symmetry known as PQ symmetry. This symmetry becomes spontaneously broken at an unknown scale and the $\theta$ angle of QCD -- which is replaced by a dynamical field in these models -- is set to zero dynamically. This solves the strong CP problem. The  dynamical field that replaces the $\theta$ angle is the axion, the light pseudo-Goldstone boson of the spontaneously broken colour anomalous symmetry \cite{Weinberg:1977ma,Wilczek:1977pj}. The axion couples only  weakly to Standard Model (SM) particles and makes for a good Dark Matter (DM) candidate \cite{Preskill:1982cy,Abbott:1982af,Dine:1982ah}. In fact, by  the so-called ``vacuum realignment'' mechanism, axions can be produced in the early universe in large enough numbers to explain the observed DM abundance. 

It has been pointed out that axion models can also provide a natural candidate for the inflaton field \cite{Fairbairn:2014zta}, thereby explaining another unsolved problem of the SM of particle physics. A recent proposal called SMASH \cite{Ballesteros:2016euj,Ballesteros:2016xej} (Standard Model-Axion-Seesaw-Higgs portal inflation) has shown how a simple axion model can be combined with the introduction of heavy right-handed neutrinos and thereby also explain both neutrino masses and the baryon asymmetry of the universe, as well as give a mechanism for inflation. The SMASH model thus addresses five fundamental problems of  particle physics and cosmology. An important feature of the model is that the scale of PQ breaking  is the only new fundamental scale of the theory and determines the masses of the heavy particles up to the effects from dimensionless couplings; for example the right handed neutrinos acquire masses exclusively from Yukawa couplings to the PQ breaking scalar.

As in most axion models, the PQ breaking scale that determines the axion mass is a free parameter. In order to constrain it one can take hints from axion cosmology and the requirement of obtaining the observed relic abundance from the realignment mechanism. The latter relies on the fact that after the phase transition in which the PQ symmetry is broken  by a vacuum expectation value (VEV) $v_{\rm PQ}$ of a complex scalar, the phase of the latter (which is proportional to the axion field) can in principle assume any initial value or ``misalignment angle'' $\theta_i$ between 0 and $2\pi$.  One can show that after the axion field gets a mass, its oscillations around the minimum contribute to the stress-energy tensor of the Universe as pressureless DM --an axion condensate-- and the observed number density of axions today depends on the value of the initial misalignment angle and on the mass of the axion at zero temperature. 
Two different scenarios can {then} be considered: (1) {If the latest phase transition that breaks the} PQ symmetry {happens} after inflation, {then} the initial misalignment angle  {is expected} to be randomly distributed throughout our observable universe and can {be} replace{d} by its average. In this case, a preferred range for the mass of the axion can be obtained. Including axion production from other effects {beyond the misalignment-induced oscillations of the axion field}, the authors of Ref. \cite{Borsanyi:2016ksw} predict an axion mass in the range $50 \,\mu {\rm eV}< m_A<  1.5\,{\rm meV}$. This scenario is usually referred to as the ``post-inflationary'' PQ breaking scenario. An alternative is given by (2), the ``pre-inflationary'' scenario{, in which} the {latest state of} PQ breaking happens already {before or} during inflation. In this case, {despite a possible initial random distribution of $\theta_i$ in patches of the size of the Hubble scale during PQ breaking, the exponential expansion of the Universe implies that our current Hubble patch can be traced to a region with a uniform value of $\theta_i$} . In this scenario -- also named the ``anthropic window'' -- one can tune $\theta_i$  to accommodate any axion mass, {and} isocurvature perturbations may place a bound on the scale of inflation. The admissible ranges for the axion mass in both scenarios are indicated in Fig.  \ref{fig:summaryshort} in lines 5 to 7.

Since the limits on the axion mass coming from cosmology are strongly dependent on the inflationary model {and the post-inflationary thermal history}, we turn to theory in the hope of obtaining predictions on the axion mass.
We have considered extensions of axion models in which the PQ solution is combined with Grand Unified Theories (GUTs) in order to obtain theoretical constraints on the axion mass. GUTs are attractive as they provide an elegant way to explain the seemingly {ad-hoc} SM gauge structure involving a product of three groups $SU(3)_C\times SU(2)_L \times U(1)_Y$. 
In unified models, the SM gauge group is embedded {into} a larger group which is {spontaneously} 
broken down to the SM at a certain 
high scale.  

The GUT proposal entails the requirement of gauge coupling unification - i.e.~the condition that the running couplings meet in the combinations defined by the GUT symmetry. 
This is a rather strict requirement which excludes the most minimal $SU(5)$ models \cite{Georgi:1974sy}. 
In models with additional particles and/or additional symmetry breaking scales, the requirement of gauge coupling unification can be implemented successfully. 
In Refs. \cite{Ernst:2018bib,DiLuzio:2018gqe}, 
we have considered  
a class of non-supersymmetric $SO(10)$ and $SU(5)$ models, endowed with an 
additional global $U(1)_{\rm PQ}$ 
symmetry that commutes with the GUT group\footnote{Note that we are not touching 
two fundamental issues 
related to such constructions, namely the gauge hierarchy between the electroweak and the GUT scale 
and the origin/quality of the global $U(1)_{\rm PQ}$. 
It is not unconceivable that a light Higgs might be dynamically selected during the evolution of the universe \cite{Dvali:2003br,Dvali:2004tma,Graham:2015cka}, 
while relaxing the simplifying hypothesis that the PQ symmetry commutes with the GUT group 
might offer new model-building directions to address the second problem \cite{Georgi:1981pu}.}. The choices of $SU(5)$ and $SO(10)$ can be motivated by minimality either of the unified gauge group --which favours the rank four $SU(5)$ over the rank five $SO(10)$-- or by minimality of the matter representations, which favour instead $SO(10)$, for which a full generation of SM particles plus a right-handed neutrino can be embedded into a single spinorial $\bf 16$ representation of $SO(10)$, which is automatically anomaly free; this is to be contrasted with the $SU(5)$ case, in which a SM generation fits into 2 representations, $\bar{\bf 5}$ and $\bf 10$, and one needs additional representations to generate neutrino masses.

Although GUT theories extended with a PQ symmetry have been considered before in the literature, our analysis offers improvements in the systematic identification of
the physical axion field, which must remain orthogonal to all the heavy gauge bosons, and the determination of its properties such as couplings to nucleons and domain-wall number. We have also performed detailed analyses of the constraints coming from unification, with the aim of studying whether they can have an impact on the allowed values of the scale of PQ breaking and the axion mass. Further constraints considered are those from proton decay, which is unavoidable in GUT theories due to the fact that unifying quarks and leptons into grand-unified representations implies an explicit breaking of the accidental baryon number symmetry of the Standard Model\footnote{Current experiments  limit the lifetime $\tau=1/\Gamma$ of the proton in the channel $p\rightarrow \pi^0 e^+$ to larger than $1.6 \times 10^{34}$ years \cite{Miura:2016krn}.}. Other bounds 
accounted for are related to axion-induced black hole superradiance and stellar-cooling effects.

The {final} goal of our analysis is the identification of a grand unified version of Refs. \cite{Ballesteros:2016euj,Ballesteros:2016xej} -- a GUT SMASH. Such a model should preferably also be able to solve the mentioned problems of the SM and modern cosmology -- neutrino masses and mixings, the strong CP problem, DM, baryogenesis and inflation.

\section{Axions in $SO(10) \times U(1)_{\rm PQ}$ theories}

\subsection{The role of the PQ symmetry}

 As explained in the introduction, the inclusion of the SM fermion representations is quite  {straightforward} in $SO(10)$ GUT models. The next step in the model building process leaves more room for speculation: the definition of the scalar sector of the theory. A hint on the possible choices can be taken from considering the Yukawa {interactions}. {Due to gauge invariance, fermion mass terms must come}  from Yukawa couplings to scalars. {To identify the allowed Yukawa interactions}, we consider the tensor product {of two fermionic $\bf 16$ representations}:
\begin{align}
 \label{eq:16times16}
\rep{16}\times \rep{16}= \rep{10}+\rep{120}+\rep{126}.
%\rep{16}_F\times \rep{16}_F= \rep{10}_H+\rep{120}_H+\rep{126}_H.
 \end{align}
 The most general Yukawa couplings can therefore be constructed using scalar fields in the representations $\rep{10}$, $\rep{120}$ and $\brep{126}$. 
For  reasons of minimality however{ we restrict ourselves} 
to only two distinct Higgs representations\footnote{A single Yukawa matrix can always be diagonalized by rotating the fermion fields, therefore {in order to allow for the observed  mixings of the SM fermions} we have to employ at least two distinct Higgs representations.}. We adhere to the most studied option and choose to employ scalar fields in the $\rep{10}_H$ and the $\brep{126}_H$ representations. As pointed out in \cite{Babu:1992ia,Bajc:2005zf}, the $\rep{10}_H$ must be taken to be complex, since otherwise the phenomenologically unacceptable mass relations $m_t\sim m_b$ are predicted. With this choice of scalar sector -- i.e.~a complex $\rep{10}_H$ and a real $\rep{126}_H$ the $SO(10)$ symmetric Yukawa Lagrangian is given by
 
 \begin{equation}
\label{eq:SO10Yukawacomplex}
\mathcal{L}_Y = \rep{16}_F \left( Y_{10} \rep{10}_H + \tilde{Y}_{10} \rep{10}_H^\ast  + Y_{126} \brep{126}_H \right) \rep{16}_F + {\rm h.c.}.
\end{equation}

After assigning vacuum expectation values (VEVs) to the included scalars in a meaningful way -- i.e.~such that the electroweak symmetry breaking can be reproduced and such that the neutrino masses are {generated} by the see-saw mechanism -- one can obtain formulae relating the fermion masses and mixings to the Yukawa matrices and VEVs. The predictive power of the model is weak, since it still contains three different Yukawa couplings (cf.~\eq{eq:SO10Yukawacomplex}). This motivated the authors of Ref.~\cite{Bajc:2005zf} to 
impose a PQ symmetry,
under which the fields transform as 
\begin{align}
  \rep{16}_F&\rightarrow \rep{16}_F e^{i\alpha}, \nonumber\\
\label{pgsymmetry_yukawas}
  \rep{10}_H&\rightarrow \rep{10}_H e^{-2i\alpha},\\
  \brep{126}_H&\rightarrow\brep{126}_H e^{-2i\alpha}\,,
\nonumber
\end{align} 
which forbids the coupling 
$\tilde{Y}_{10}$ in \eqref{eq:SO10Yukawacomplex} (see also Ref.~\cite{Babu:1992ia}).
This is how in many GUT models, a PQ {symmetry} is {invoked} without {a} reference to the strong CP problem!

\subsection{Definition of the models}

All $SO(10)$ models considered here have in common that they employ fermionic representations in the $\rep{16}_F$ and scalar representations in the $\rep{10}_H$ and the $\brep{126}_H$, which transform as given in \eqref{pgsymmetry_yukawas} under a global PQ symmetry. This scalar particle content however is not sufficient to break the GUT group down to the SM gauge group, and we must employ at least one additional scalar field. 
We choose\footnote{{A more minimal option would be given by a (real) $\rep{45}_H$. 
This representation was doomed in the early 80's \cite{Yasue:1980fy,Anastaze:1983zk,Babu:1984mz}}
{due to the emergence of tachyons  in the tree-level scalar spectrum, 
arising for the phenomenologically relevant non-$SU(5)$-like breaking patterns 
required by gauge coupling unification \cite{Deshpande:1992em,Bertolini:2009qj}.} 
{However, it was recently shown \cite{Bertolini:2009es,Bertolini:2012im,Bertolini:2013vta,Graf:2016znk}} 
{that the tachyonic instabilities can be removed at the one-loop level, 
thus re-opening for exploration the most minimal $SO(10)$ Higgs sector. 
}} 
the $\rep{210}_H$ \cite{Ernst:2018bib}, which contains a singlet under the Pati-Salam group 
$SU(4)_C\times SU(2)_L\times SU(2)_R$, such that our symmetry breaking chain is
\begin{eqnarray} 
\label{chaintwostep}
SO(10)&\stackrel{v^{210}-210_H}{\longrightarrow}
&4_{C}\, 2_{L}\, 2_{R}\ \stackrel{v_R -\overline{126}_H}{\longrightarrow}3_{C}\, 2_{L}\, 1_{Y}\ \stackrel{v_{u,d}^{10,126}-10_H}{\longrightarrow} \ 3_{C}\,1_{\rm em}\,.
\end{eqnarray}
The symmetry breaking VEVs indicated above the arrows are constrained by the requirement of gauge coupling unification. 

We have not fixed yet  the PQ charge of this additional scalar. The most minimal option might be to leave the $210_H$ a PQ singlet. A more careful analysis of such a model (see also \cite{Mohapatra:1982tc}) results however
in an axion decay constant at the electroweak scale -- i.e.~the resulting model contains a visible axion which is already excluded by experiments. We must therefore construct models in which the axion decay constant is lifted from the electroweak scale. This can be achieved by {ensuring that more than one of the scalar multiplets acquiring large VEVs are charged under PQ, for example by} (1) extending the PQ symmetry to all existing multiplets, (2) postulating an additional multiplet  and extending the PQ symmetry to it or (3) postulating an additional singlet and extending the PQ symmetry to include it. We consider models of all three types, where the additional multiplet in models of the second type transforms as a $\rep{45}_H$ representation. The PQ charges of these models are summarized in table \ref{tab:PQassignments}. The original versions of these models are variations of the original DFSZ \cite{Dine:1982ah,Zhitnitsky:1980tq} axion model and have a non-trivial domain wall number $N_{\rm DW}$
of 3.\footnote{In Ref. \cite{Ernst:2018bib} we clarify how the domain wall number corresponds to the dimension of the finite group that arises from the translational symmetry of the axion-Goldstone after the explicit breaking of the anomaly, and after modding out by unphysical rotations that leave the scalar fields invariant, as well as by transformations in the center of the gauge group. We also show how the domain-wall number can be computed entirely in terms of the charges of the scalars for the physical PQ symmetry that gives an axion that does not mix with the heavy gauge bosons .} 
The table also indicates variations of these models in which the domain wall number is reduced to 1 by including exotic fermions. 

Let us briefly summarize the defining properties of each model before moving on to the computed constraints:
\paragraph{Model 1}
The $\rep{210}_H$ -- being the only  multiplet {that had no charge under} $U(1)_{\rm PQ}$ -- is assigned a PQ charge. Its VEV then sets the scale of the axion decay constant, which is inversely proportional to the axion mass. {Moreover,} the VEV of the $\rep{210}_H$  is also {the one} that breaks the GUT symmetry  {and} is therefore constrained by the requirement of gauge coupling unification and {by} proton decay {searches}. {From the above it follows that} this model features an axion decay constant at the unification scale. 
\paragraph{Models 2.1 and 2.2}
Both models feature an additional multiplet in the $\rep{45}_H$. This choice is interesting as it allows for an additional step in the symmetry breaking chain:
\begin{eqnarray}
\label{chain2}
SO(10)&\stackrel{M_{\rm U}-210_H}{\longrightarrow}
4_{C}\, 2_{L}\, 2_{R}\, \stackrel{M_{\rm PQ}-4{\bf 5}_H}{\longrightarrow}
4_{C}\, 2_{L}\, 1_{R}\, \stackrel{M_{\rm BL}-126_H}{\longrightarrow}
3_{C}\, 2_{L}\, 1_{Y}\, \stackrel{M_Z-10_H}{\longrightarrow} \ 3_{C}\,1_{\rm em}.
\end{eqnarray}
The scales of symmetry breaking and {the} responsible multiplets are indicated above the arrows.
This breaking chain occurs only if $M_{\rm PQ}> M_{\rm BL}$, otherwise the relevant breaking chain is given by \eqref{chaintwostep}. The axion decay constant is given by an intermediate scale $M_{\rm PQ}$ which is constrained by the gauge coupling unification requirement. 

Model 2.2 also features two exotic fermion representations in the $\rep{10}_F$, {which ensure a domain-wall number equal to one}.
\paragraph{Models 3.1 and 3.2}
The least constraining extension of our original model is {obtained} by including a gauge singlet charged under the PQ symmetry. This singlet is then not constrained by gauge coupling unification and can take any VEV. In these models the axion decay constant is not constrained by theoretical considerations. Model 3.2 is a variation featuring two exotic fermion representations in the $\rep{10}_F$, {again giving $N_{DW}=1$}.

\begin{table}[h]
\begin{align*}
\begin{array}{ l c c c c c c c c }
\toprule
 & \rep{16}_F & \brep{126}_H & \rep{10}_H & \rep{210}_H  & \rep{45}_H & S & \rep{10}_F & N_{\rm DW} \\
\midrule
{\bf Model\ 1}  & 1 & -2 &-2 & 4  & - & -& -  & 3  \\
\midrule
{\bf Model\ 2.1}  & 1 & -2 &-2 & 0  & 4 & -& -  & 3   \\
\midrule
{\bf Model\ 2.2}  & 1 & -2 &-2 & 0  & 4 & -& -2 & 1   \\
\midrule
{\bf Model\ 3.1}  & 1 & -2 &-2 & 0  & - & 4 & -  & 3 \\
\midrule
{\bf Model\ 3.2}  & 1 & -2 &-2 & 0  & - & 4 & -2  & 1
\\
\bottomrule
\end{array}
 \end{align*}
 \caption{\label{tab:PQassignments}Field content, PQ charge assignments, and resulting domain wall number $N_{\rm DW}$ in the various $SO(10)\times U(1)_{\rm PQ}$ models considered in{ \cite{Ernst:2018bib}}.}
\end{table}

\subsection{Results}

We have analyzed each of the above model for the constraints put on the axion mass by the requirement of gauge coupling unification and experimental limits \cite{Ernst:2018bib}. We have made a two-loop analysis, which requires the inclusion of one-loop threshold corrections. The threshold corrections depend strongly on the masses of the heavy scalars that need to be integrated out at their corresponding scales. In lack of a detailed knowledge of the rather involved scalar sector we have 
{assumed the so-called ``extended survival hypothesis'' \cite{delAguila:1980qag,Mohapatra:1982aq} } -- yet modified to allow for two light Higgs doublets at low scales -- and
scanned over randomized values for the scalar masses. These masses are assumed to be in the range $[\frac{1}{10\,} M_T, 10 \,M_T]$, where $M_T$ is a threshold, i.e.~a symmetry breaking scale. The {possibility of}  rather large threshold corrections  {is} the reason for the {wide axion mass ranges}  predicted in our analysis.

\begin{figure}[h]
\begin{centering}
\includegraphics[width=\textwidth]{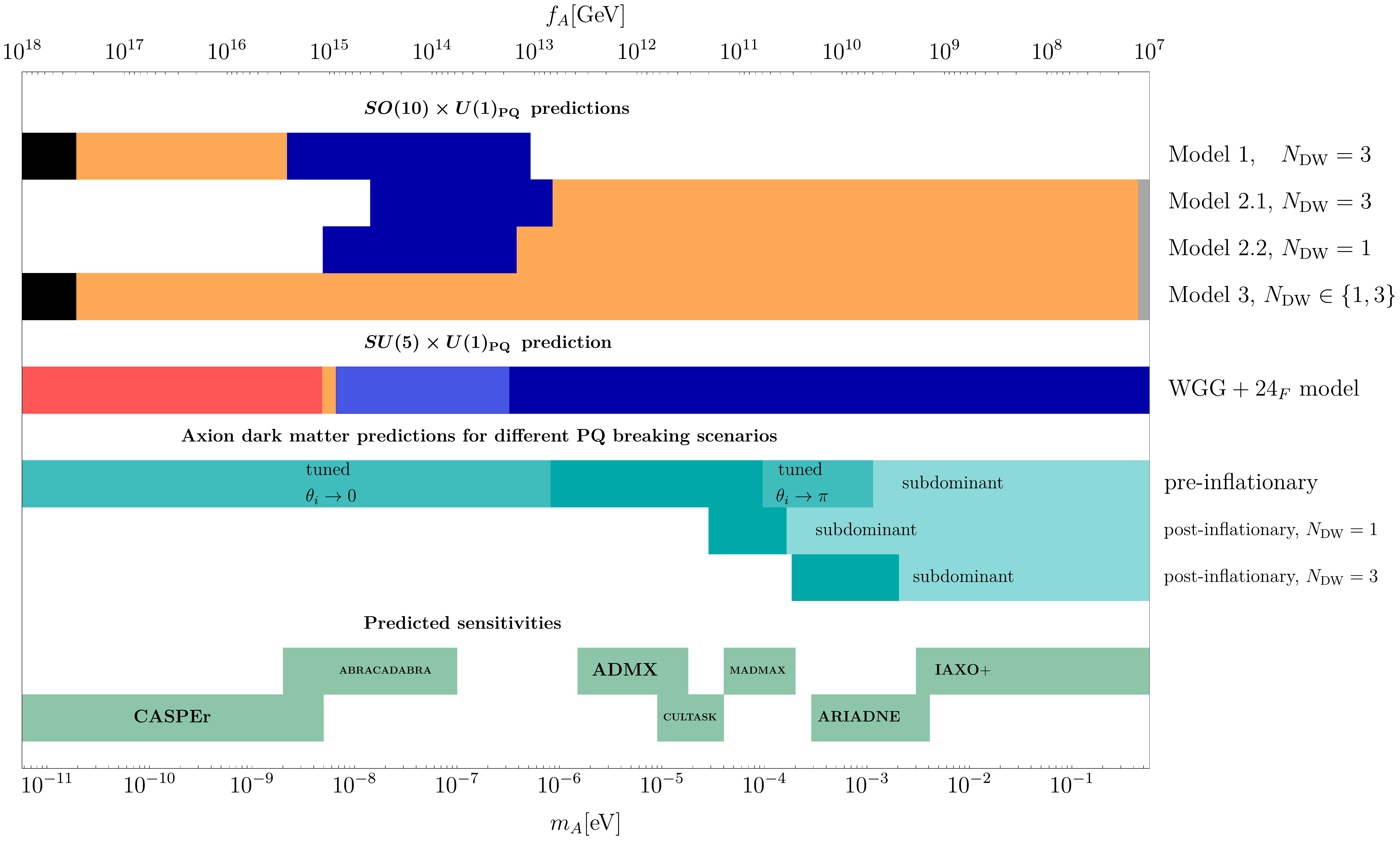}\\
\end{centering}
  \caption{\label{fig:summaryshort} 
Possible ranges of the axion mass and decay constant consistent with gauge coupling unification in our four models.
Regions in black are excluded by constraints from black hole superradiance \cite{Arvanitaki:2014wva}, 
regions in dark blue by proton stability constraints, and areas in red by LHC constraints. 
We have also included the much more constrained prediction for the axion mass range in a minimal $SU(5)\times U(1)_{\rm PQ}$ model described in section \ref{subsec:SU5prediction} \cite{DiLuzio:2018gqe}, denoting in light blue the region in which the proton decay bounds can be circumvented with appropriate tunings.
Regions in grey are excluded by stellar cooling constraints from horizontal branch stars in globular clusters  \cite{Ayala:2014pea}. For comparison, we show also the mass regions preferred by axion DM (lines 6 to 8), cf.~\cite{Saikawa:2017lzn}. Here, the dark regions indicate the ranges where the axion can make up the main part of the observed DM with a tuned misalignment angle in the pre-inflationary PQ breaking scenario. In the light regions, axions could still be DM, but not the dominant part. The remaining regions are not allowed - axions in this mass range would be overabundant. 
Note that the region in the $N_{\rm DW}=3$ case has been derived under the assumption that the PQ symmetry is protected by a discrete symmetry, so that Planck scale suppressed PQ violating operators are allowed at dimension 10 or higher \cite{Ringwald:2015dsf}. 
In the last two lines the projected sensitivities of various experiments are indicated \cite{Budker:2013hfa,Stern:2016bbw, Chung:2016ysi,Kahn:2016aff,TheMADMAXWorkingGroup:2016hpc, Arvanitaki:2014dfa,Armengaud:2014gea,Giannotti:2017hny}. A similar plot was already published in Ref. \cite{Ernst:2018bib}. }
\end{figure}

We have summarized our results in Fig. \ref{fig:summaryshort}. The allowed ranges of the axion mass and decay constant are plotted in orange for each of our models. We also compare our results to the reaches of various axion experiment and also to the permitted mass ranges in the various inflationary scenarios. 
\paragraph{Model 1}
The most constraining -- and also the most minimal -- of our $SO(10)$ models is Model 1. 
The axion mass is suppressed by the GUT scale, and we predict an axion in the range
\begin{equation}
1.9 \times 10^{-11}\eV < m_A<2.2\times 10^{-9} \eV.
\end{equation} 
As indicated in Fig. \ref{fig:summaryshort}, this model only allows for the pre-inflationary PQ breaking scenario, 
{since the PQ and GUT breaking scale are tightly connected.} 
In order for axion cold DM not to become overabundant, the initial value of the axion field 
in the causally connected patch which contains the present universe had to be small, 
$10^{-3}\lesssim |\theta_i| = |A(t_i)/f_A|\lesssim 10^{-2}$  \cite{Borsanyi:2016ksw}. 

Remarkably, the predicted axion mass range of Model 1 will be probed in the 
next decade by the CASPEr-Electric experiment \cite{Budker:2013hfa}, cf.~Fig. \ref{fig:CASPErelectric}, 
which aims to probe the axion-induced electric
dipole moment (EDM) of the nucleon,
$d_N (t) = g_{d}\,\frac{\sqrt{2\rho_{\text{DM}}}}{m_A} \cos (m_A\,t)$ \cite{Graham:2013gfa}, {
where $g_{d}$ is the model-independent coupling of the axion to the  EDM operator $\mathcal{L}_A \supset -\frac{i}{2} g_{d}\, A\, \overline{\Psi}_N \sigma_{\mu\nu} \gamma_5 \Psi_N F^{\mu\nu}$ of the nucleon
and $\rho_{\rm DM}=0.3\,\text{GeV}/\text{cm}^3$ is the local density of axion DM.
If successful and interpreted in terms of Model 1, one may translate the measurement of the axion mass into an indirect determination of the mass of the heaviest gauge bosons, i.e.~the unification scale, 
\begin{align}
\label{eq:MU_determination_model1}
 \MU \simeq   3 g_U \sqrt{\chi}/m_A ,
\end{align}
where $\chi$ is the topological susceptibility in QCD, $\chi = [75.6(1.8)(0.9) {\rm MeV}]^4$ \cite{diCortona:2015ldu,Borsanyi:2016ksw} and $g_U$ is the unified gauge coupling at the unification scale. 
The unification scale 
can be probed complementarily by 
the next generation of experiments looking for signatures of proton decay, such as Hyper-Kamiokande \cite{Abe:2011ts} or DUNE \cite{Kemp:2017kbm}. 

\begin{figure}[h]
 \begin{centering}
 \hspace{50pt}
  \includegraphics[width=0.7\textwidth]{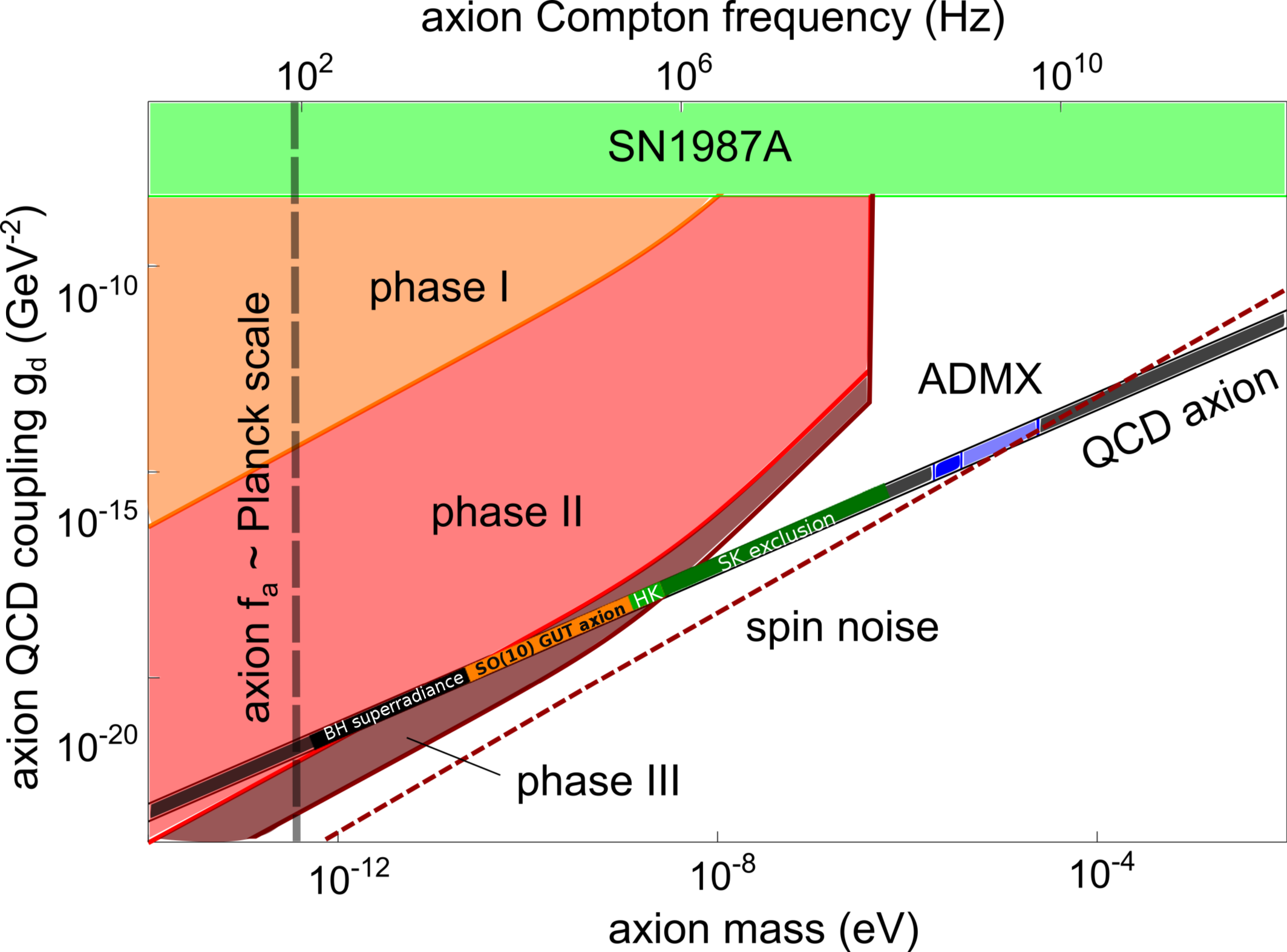}
 \end{centering}
\caption{\label{fig:CASPErelectric} 
Plot adapted from Ref. \cite{Budker:2013hfa}.
Experimental reach of CASPEr-Electric. The QCD axion is indicated by a band in the parameter space, and the range predicted by Model 1 is indicated in orange and light green. Black hole superradiance limits are drawn {again} in black, and the range excluded by proton decay limits is indicated in dark green, to avoid confusion with the ADMX sensitivity range painted in blue. The small light green region labeled HK indicates the predicted range  {within the sensitivity of} the Hyper-Kamiokande experiment {after 10 years of data collection}. {The orange band gives the predicted region outside the reach of Hyper-Kamiokande.}  The light orange, red, and maroon regions demonstrate the predicted sensitivity of the CASPEr-Electric experiment in the phases I-III as explained in Ref. \cite{Budker:2013hfa}. Phase III will be able to reach the $SO(10)$ - GUT axion as predicted by our Model 1.
}
\end{figure}
\paragraph{Other models}
In Models 2.1 and 2.2, the requirement of gauge coupling unification constrains the intermediate scale which sets the axion decay constant in these models if the PQ breaking is correlated with the breaking of a gauge group. This however is only the case if the PQ breaking scale is larger than the scale at which the B-L symmetry is broken. Thus, no lower limit on $f_A$ can be derived in these models. An upper bound is given by $f_A<1.5 \times 10^{13} {\rm GeV}$. In \mbox{Model 3}, no constraints on the axion decay constant can be derived.

\section{A predictive $SU(5) \times U(1)_{\rm PQ}$ model}
\label{sec:SU5}

Motivated by the quest for a more predictive axion-GUT framework,  
we turn to $SU(5)$ grand unification.
The simplest {extension of}  non-supersymmetric $SU(5)$ {to accommodate the axion} was proposed by 
Wise, Georgi and Glashow (WGG) \cite{Wise:1981ry} (see also \cite{Nilles:1981py}). 
{Nevertheless}, the WGG model 
is ruled out because of unsuccessful gauge coupling unification and zero neutrino masses, {much like}  the original $SU(5)$ model of Georgi and Glashow (GG) \cite{Georgi:1974sy}.  
A solution to these issues in the GG model was proposed in Refs.~\cite{Bajc:2006ia,Bajc:2007zf}, {relying on extending the field content with one  fermion
multiplet ${\bf 24}_F$ in the adjoint of $SU(5)$. The latter allows to give rise to neutrino masses through a combined Type-I+III seesaw mechanism, and 
can also lead to successful unification. A thorough renormalization group analysis}  \cite{Bajc:2006ia,Bajc:2007zf,DiLuzio:2013dda} {leads to
the prediction of a sharp correlation between the masses of  light electroweak triplets (which may be probed at the Large Hadron Collider (LHC)) 
and the scale of unification, which is constrained by the proton-decay bounds from Super-Kamiokande (SK)).}

In Ref.~\cite{DiLuzio:2018gqe} we have extended the WGG model 
with a ${\bf 24}_F$, in analogy to the GG+${\bf 24}_F$ case\footnote{From the point of view of the GG+${\bf 24}_F$ model, 
the PQ extension is also motivated by the absence of a DM candidate.}.
Next we will give a succinct review of the WWG model and its +${\bf 24}_F$ extension, 
focusing in particular on the connection of the axion mass with the GUT scale.  
Combining constraints from proton stability, LHC and unification 
we obtain the axion mass prediction: $m_A \in [4.8, 6.6 (330)]$ neV, 
where the number in parenthesis corresponds to the case of
a tuned flavour structure in the operators that mediate proton decay.

\subsection{The Wise-Georgi-Glashow model} 
{Here we review the main characteristics of the WGG model} \cite{Wise:1981ry}. {As in the original GG model} \cite{Georgi:1974sy}, {the 
SM fermions fit into three copies of $\bar {\bf 5}_F$ and ${\bf 10}_F$, and the scalar sector includes a complex ${\bf 24}_H$ and two fundamentals, ${\bf 5}_H$ and ${\bf 5}'_H$.
The Yukawa Lagrangian is\footnote{{It should be noted that obtaining the correct ratio between the masses of down quarks and charged leptons requires
the addition of nonrenormalizable operators or new scalar representations.}}}
\beq 
\label{Lyuk}
\mathcal{L}_{Y} = \bar {\bf 5}_F {\bf 10}_F 5'^*_H + {\bf 10}_F {\bf 10}_F {\bf 5}_H \, ,
\eeq
{whereas the scalar potential contains the following interactions transforming nontrivially under global rephasings,}
\beq 
V_H \supset{\bf 5}'^\dag_H {\bf 24}^2_H {\bf 5}_H +{\bf 5}'^\dag_H {\bf 5}_H \Tr ({\bf 24}^2_H) \, ,
\eeq
{which are similar to those in DFSZ models}  \cite{Zhitnitsky:1980tq,Dine:1981rt}. {It turns out that the terms in the previous equation, and the entire Lagrangian, are invariant under the following $U(1)_{\rm PQ}$ symmetry:}
$\bar {\bf 5}_F \to e^{-i\alpha/2} \bar {\bf 5}_F$, ${\bf 10}_F \to e^{-i\alpha/2} {\bf 10}_F$, 
${\bf 5}_H \to e^{i\alpha} {\bf 5}_H$, ${\bf 5}'_H \to e^{-i\alpha} {\bf 5}'_H$ and ${\bf 24}_H \to e^{-i\alpha} {\bf 24}_H$.     

{The GUT gauge group is broken down to the SM group by means of a VEV in the ${\bf 24}_H$. Representing the adjoint field by a $5\times5$ matrix,
the appropriate VEV, the associated axion excitation and the axion decay constant $f_A$ are given by
\beq 
\vev{{\bf 24}_H} = V \tfrac{1}{\sqrt{30}} \, \text{diag} (2,2,2,-3,-3), \quad {\bf 24}_H \supset \vev{{\bf 24}_H} \tfrac{1}{\sqrt{2}} e^{i \, A / V}, \quad f_A = \frac{V}{\hat{N}},
\eeq
with $\hat{N}$ the anomaly coefficient of the $U(1)_{\rm PQ}$ symmetry under $SU(3)_C$ -- equal to 6 in the WGG+${\bf 24}_F$ model.
$V$ fixes the mass of the heavy leptoquark vector bosons, which are also proportional to the $SU(5)$ gauge coupling $g_5$:
\beq 
\label{mVWGG}
m_V = \sqrt{\tfrac{5}{6}} g_5 V = \sqrt{\tfrac{5}{6}} \hat{N} g_5 f_A \,.
\eeq
Crucially, the heavy gauge boson mass, which is constrained by proton decay experiments, is connected to the axion decay constant, 
which is delimited by axion searches because it fixes the axion mass. Both types of experiments are complementary probes of 
the model and there is a direct connection between the axion mass 
and the proton decay rate. To see this connection explicitly we may start from the following formula of the proton decay rate in the 
$p \to \pi^0 e^+$ channel, obtained within chiral perturbation theory and recasted for $SU(5)$} \cite{Claudson:1981gh,Nath:2006ut}: 
\begin{equation}
\label{Gammapdecay1}
\begin{aligned}
&\Gamma_{p \to \pi^0 e^+} =\, \frac{m_p}{16 \pi f_\pi^2} A^2_L \abs{\alpha}^2 (1+D+F)^2 
\left(\frac{g_5^2}{2 m^2_V}\right)^2 
\left[ 4 A^2_{SL}  + A^2_{SR} \right],\\
%%%%
&\begin{array}{c c c}
m_p=   938.3 \text{ MeV},&  f_\pi = 139 \text{ MeV},& \alpha = -0.011 \text{ GeV}^3,\\
%%%%
A_L=1.25,& D=0.81, & F=0.44 .
%%%
\end{array}
\end{aligned}
\end{equation}
{
The previous formula ignores fermion mixing effects. $A_L$ includes effects from renormalization between the proton mass and the electroweak scales, 
while $A_{SL(R)}$ are renormalization factors from the electroweak to the GUT scale } \cite{Buras:1977yy,Wilczek:1979hc,Bertolini:2013vta}.{ These depend
on intermediate thresholds, and for SM running up to $10^{15}$ GeV one has $A_{SL(R)} = 2.4 \, (2.2)$. The rest of the constants in} \eqref{Gammapdecay1} {are phenomenological parameters of the chiral Lagrangian. Substituting the gauge boson mass \eqref{mVWGG} and using the relation
$m_A = 5.7 \, \text{neV} \, (10^{15} \, \text{GeV} / f_A)$ }\cite{diCortona:2015ldu,Borsanyi:2016ksw} one obtains:
\begin{equation}
\label{Gammapdecay2}
\Gamma_{p \to \pi^0 e^+} \simeq \left( 1.6 \times 10^{34} \ \text{yr} \right)^{-1} \left( \frac{m_A}{3.7 \ \text{neV}} \right)^4 
\left( \frac{6}{\hat{N}} \right)^4
\left[ 0.83 \left( \frac{A_{SL}}{2.4} \right)^2 + 0.17 \left( \frac{A_{SR}}{2.2} \right)^2 \right] 
\, , 
\end{equation}
where the first factor is 
the current proton decay bound from SK \cite{Miura:2016krn}.  

\subsection{Axion mass prediction in Wise-Georgi-Glashow$+{\bf 24}_F$ model}
 \label{subsec:SU5prediction}

 {In the presence of the additional fermion representation in the ${\bf 24}_F$, the 
Yukawa Lagrangian is augmented by 
\beq 
\label{DeltaLY}
\Delta \mathcal{L}_Y = \bar {\bf 5}_F {\bf 24}_F {\bf 5}_H + 
\Tr {\bf 24}^2_F {\bf 24}^*_H  \, .
\eeq 
The first term generates Yukawa interactions for the SM triplet and singlets within the  ${\bf 24}_F$, and the 
second interaction gives  a Majorana mass for the full multiplet after the breaking of $SU(5)$. One also needs
nonrenormalizable operators to get at least two distinct light neutrino masses, and to split the masses
of the ${\bf 24}_F$ sub-multiplets} \cite{Bajc:2006ia,Bajc:2007zf,DiLuzio:2013dda}, {but for our purposes 
it suffices to note that} \eqref{DeltaLY} {fixes the PQ charge of the ${\bf 24}_F$ at $-1/2$, which in turn 
gives an anomaly coefficient of $\hat{N}=11$ for the PQ symmetry under $SU(3)_C$.

Unification constraints are rather strict, and allow to narrow down the axion mass range due to the connection 
between the axion mass and the unification scale. In the SM, unification fails because the electroweak 
couplings  $\alpha_1$ and $\alpha_2$ meet at a scale around $10^{13}$ GeV, which is ruled out by proton decay bounds.
Thus, viable unification requires matter multiplets that can delay the meeting of $\alpha_1$ and $\alpha_2$, mainly by making 
 $\alpha_2$ increase faster for a growing scale (additional multiplets with nonzero hypercharge will make $\alpha_1$ increase faster and worsen
 the problem). In the WGG+${\bf 24}_F$ model, each $\bf 24$ representation (fermion or scalar) contains the following matter multiplets of the SM group $SU(3)\times SU(2)_L\times U(1)_Y$:
 \begin{align}
  {\bf 24}=({\bf 8},{\bf 1}, 0)+({\bf 3},{\bf 2},- 5/6)+({\bf {\bar 3}},{\bf 2}, 5/6)+({\bf 1},{ \bf 3},0)+({\bf 1},{\bf 1},0).
 \end{align}
Then the appropriate saviour role can be played by the $SU(2)$ triplet $T_F = ({\bf 1},{\bf 3},0)\subset{\bf 24}_F$, and similarly 
 for the triplet scalar $T_H = ({\bf 1},{\bf 3},0)$ within the ${\bf 24}_H$. Both triplets are predicted to be near the weak scale, so as to maximize the effect on
 the running of $\alpha_2$, and could potentially be probed at the LHC. The fermion triplet can give rise to lepton-number violation, 
 same sign di-lepton events}
\cite{Arhrib:2009mz};  {current CMS bounds for $m_{T_F}$ are near $840$ GeV} \cite{Sirunyan:2017qkz}, {while
 High Luminosity LHC (HL-LHC) projections can reach up to 2 TeV }\cite{Ruiz:2015zca,Cai:2017mow}. {In turn, scalar triplets can impact the di-photon Higgs signal, 
 yet the effect is model-dependent and smaller to that of the fermionic triplet} \cite{Chabab:2018ert}.
 
 {
 Finally, requiring also the convergence  of $\alpha_3$ 
with $\alpha_1$ and $\alpha_2$ demands heavier colored particles, which can be found in the the $({\bf 8},{\bf 1},0)$ color-octet scalars and fermions 
within the ${\bf 24}_{F,H}$. Their  masses are  required to be around $10^8$ GeV, 
far from the LHC reach. 
 
 We have carried out a gauge coupling unification analysis following} \cite{DiLuzio:2013dda},
 {accounting for the 3-loop beta functions 
induced by the scalar and fermion triplets, and 
 the leading NNLO corrections 
from the 2-loop threshold effects. Including the constraints from the present LHC bounds and SK 
the resulting axion mass window is 
\beq 
\label{eq:preferred_window}
m_A \in [4.8, 6.6]\  \text{neV}\,.
\eeq
The upper bound can be relaxed when allowing for cancellations in the  flavour structure of proton decay operators} \cite{Dorsner:2004xa,Kolesova:2016ibq}, 
{which can be used to tune to zero the proton decay rate in a number of channels. Unitarity of the mixing matrices however implies that some channels will always survive,
and the weakest constraint is obtained when only decays to strange mesons are allowed} \cite{Dorsner:2004xa}. {This gives  $\tau / \mathcal{B} (p \to K^0 \mu^+) > 1.3 \times 10^{33}$ yr} \cite{Regis:2012sn}, {which implies an absolute upper bound for the axion mass of  $m_A < 330$ neV. Concerning future experiments, the projected sensitivity of HK} \cite{Abe:2014oxa}
{in the $p \to K^+ \bar \nu$ channel could rule out $m_A > 160$ neV. 

Aside from studying unification constraints and their effect on axion masses, we calculated the axion couplings to SM particles and studied the reach of axion experiments probing 
the coupling to photons or nucleons. The ABRACADABRA experiment}  \cite{Kahn:2016aff} {is posed to reach enough sensitivity to probe the axion photon coupling $g_{A\gamma}$, defined by  ${\cal L}_A \supset \frac{1}{4} g_{A\gamma}\,a\,F_{\mu\nu} {\tilde F}^{\mu\nu}$, 
in the relevant mass region. We show this in the left panel of } \fig{fig:gagg_vs_ma},{ from which we  conclude
that the late stages of the experiment can test the whole parameter space 
of the WGG+${\bf 24}_F$ model, with the tuned region included. 
In the right panel of Fig.}~\ref{fig:gagg_vs_ma} we display the projected sensitivity 
of CASPEr-Electric \cite{Budker:2013hfa,JacksonKimball:2017elr}. 
We took the non--perturbative 
estimate of the axion coupling $g_{AD}$ to the EDM operator of the nucleon from } Ref. \cite{Pospelov:1999mv} {and the QCD axion band in the figure 
shows the theoretical error. Phase III of CASPEr-Electric could probe the preferred axion mass window} \eqref{eq:preferred_window}. 
A dedicated measurement time focused on the preferred mass region could give a factor of three improvement in the reach, since  the sensitivity in $g_{AD}$ improves with the scanning time 
as $t^{1/4}$. We denote this optimal reach with  a short, full blue line in the 
right panel of \fig{fig:gagg_vs_ma}.

\begin{figure}[ht!]
\includegraphics[width=.5\textwidth]{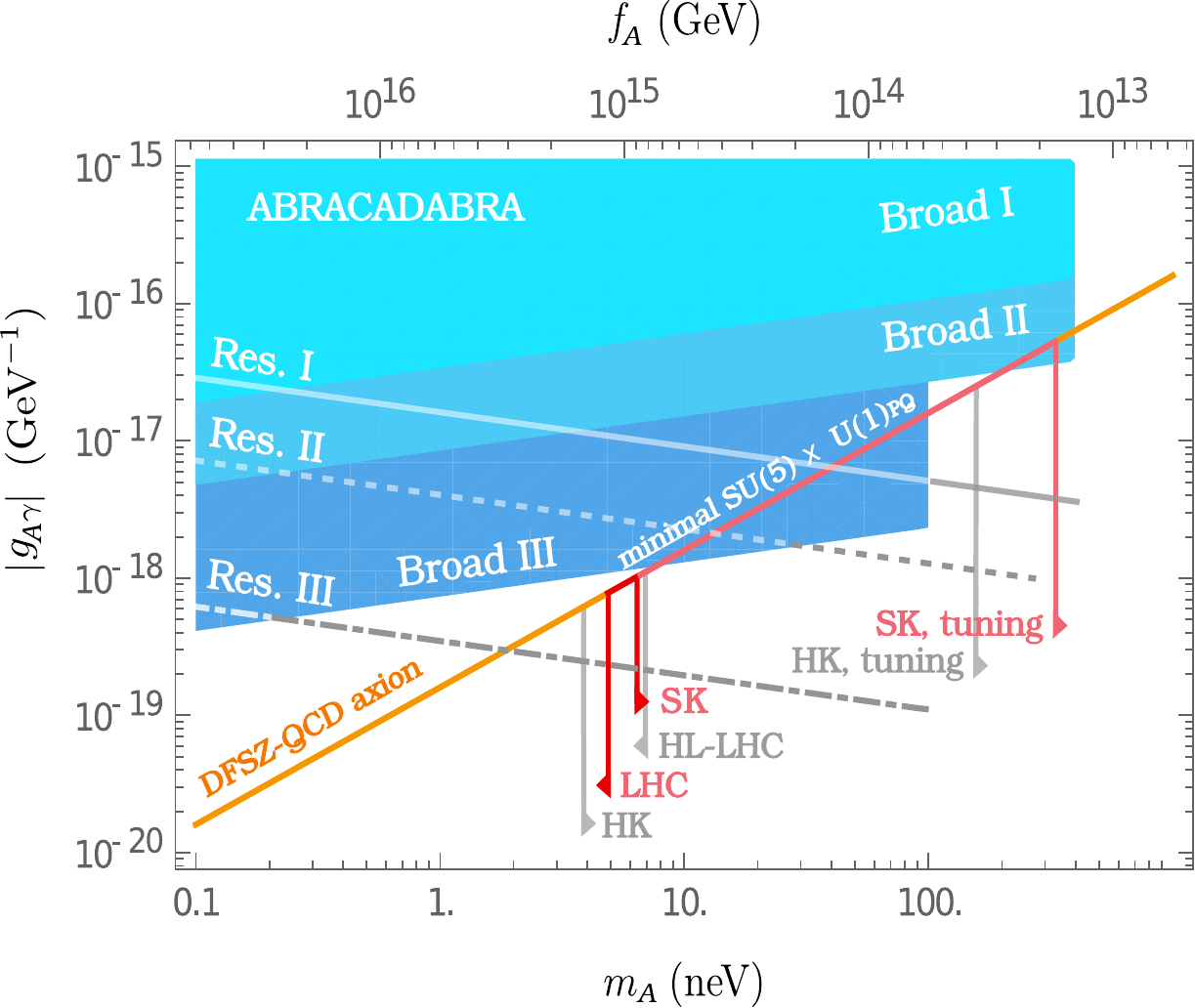}
\includegraphics[width=.5\textwidth]{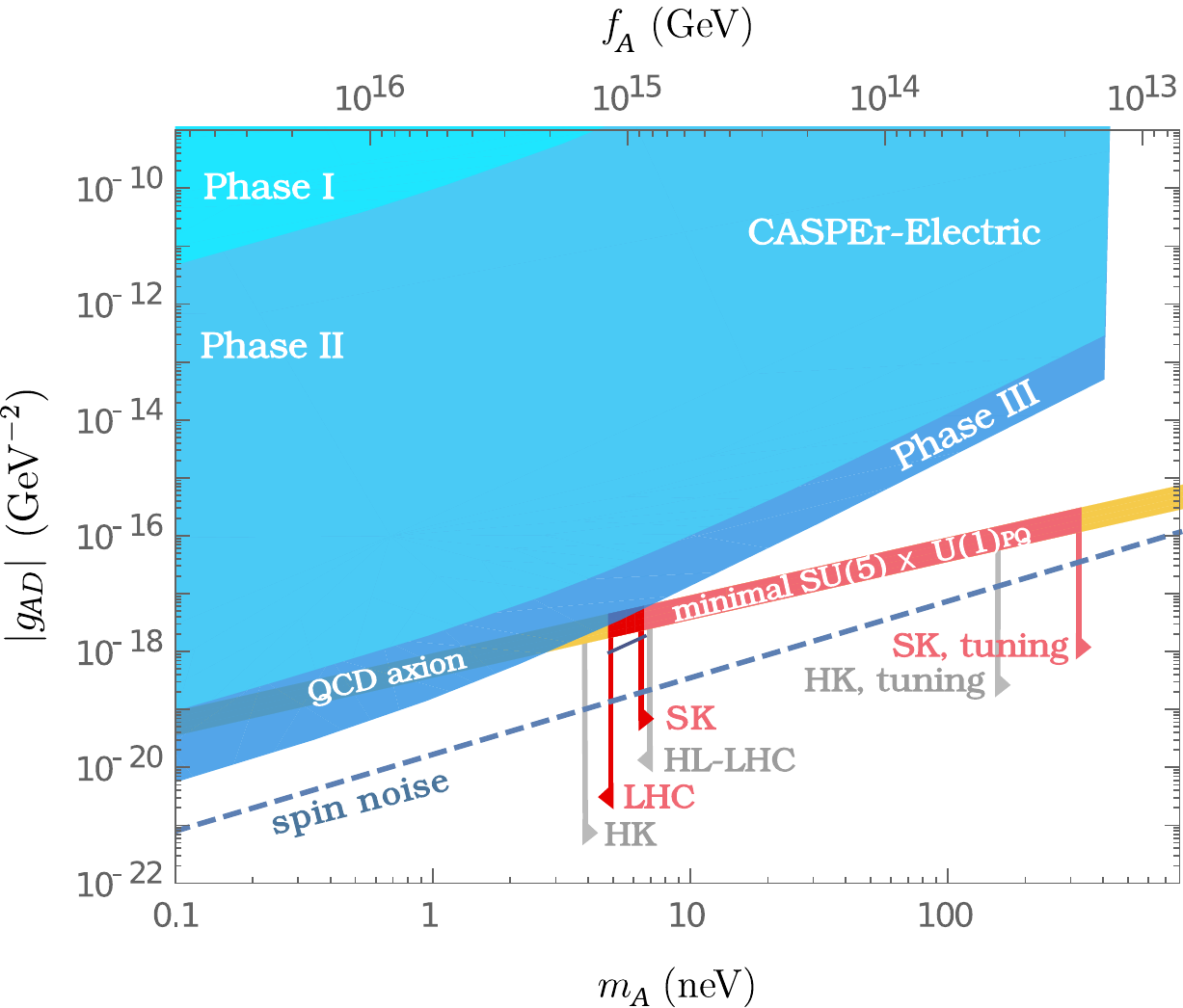}
\caption{\label{fig:gagg_vs_ma}
Left panel: Axion coupling to photons, $g_{A\gamma}$, against the axion mass $m_A$. 
The blue areas give the estimated sensitivities of the broadband (``Broad'') 
and resonant (``Res.'') search modes of ABRACADABRA, taken 
from Ref.~\cite{Kahn:2016aff}. 
Right panel: Axion coupling $g_{AD}$ to the EDM operator of the nucleon against the axion mass $m_A$. 
The blue areas show the estimated sensitivities of CASPEr-Electric, taken 
from Ref.~\cite{JacksonKimball:2017elr}. The short, full blue line corresponds to a factor of three enhancement of
the sensitivity for an optimized search focused on the favoured mass region. Both figures have been published in Ref.~\cite{DiLuzio:2018gqe}.}
\end{figure}

\section{Conclusions}

The axion is a well-motivated hypothetical particle that is not only  part of a mechanism that solves the strong CP problem, but is also able to explain the DM abundance of the universe. As a consequence, a comprehensive experimental program has developed that could probe large swaths of the parameter space predicted by the QCD axion in the coming decade. Nevertheless, this is a vast parameter space, as the PQ breaking scale that determines the axion mass and its couplings to SM particles is typically a free parameter. For these reasons it becomes interesting to look for possible theoretical constraints that may motivate specific choices of parameters for which experiments or scanning strategies could be optimized.

In our recent work \cite{Ernst:2018bib,DiLuzio:2018gqe}, we investigated whether embedding the axion into Grand Unified Theories may provide such theoretical constraints. With GUTs themselves offering a compelling --and constrained-- unified framework for the gauge structure of the SM, while at the same time allowing for consistent cosmological histories, the study of the interplay between axion physics and the requirements from unification is well motivated. In our research we studied the properties of the axion in $SO(10)$ and $SU(5)$ theories extended with an anomalous global $U(1)_{\rm PQ}$ symmetry. We calculated the axion mass in terms of the threshold scales, and obtained the axion couplings to the SM particles. A detailed renormalization group
analysis allowed us to assess the impact of the constraints for unification in the axion mass, for which we also accounted from bounds from proton decay and collider searches. 

In those $SO(10)$ and $SU(5)$ models, in which all the Higgs multiplets involved in the breaking towards the SM carried PQ charges, the PQ breaking scale and thus the axion mass is rather constrained by gauge coupling unification 
and can be probed by axion DM experiments, such as 
ABRACADABRA and CASPEr-Electric. Moreover, the relation between the GUT and PQ scale implies that the axion can be constrained as well by proton decay searches. The $SU(5)$  WGG+${\bf 24}_F$ model is particularly predictive and gives an allowed window of $m_a\in[4.8,6.6(330)]$ neV, with the upper limit depending on whether one allows for tuning in the fermion mixing matrices. Interestingly, in this case unification demands light electroweak triplets and the axion mass can also be  probed indirectly through collider searches, so that the narrow mass window could be fully explored by both HL-LHC and HK, aside from the direct DM searches.

\bibliographystyle{unsrt}
\bibliography{bibfile}

\end{document}